# Mapping of Deformation to Apparent Young's Modulus in Real-Time Deformability Cytometry

Christoph Herold

ZELLMECHANIK DRESDEN GmbH · Tatzberg 47/49 · 01307 Dresden · Germany
herold@zellmechanik.com

As described in the work of Mietke *et al.* (*1*) the deformation (defined as 1 – circularity [see (*2*)]) of a purely elastic, spherical object deformed in a real-time deformability cytometry (RT-DC) experiment can be mapped to its apparent Young's Modulus.

This note is supposed to help a fast and correct mapping of RT-DC results – namely, deformation and size – to values of the apparent Young's Modulus *E*.

In experiments with a controlled constant flow rate, the absolute values of *E* obtained by mapping according to (*1*) depend on the exact knowledge of the channel cross-section and the viscosity of the surrounding medium. Another potential caveat for correct mapping rises from the pixelation of the image and the influence thereof on the deformation and size value calculated from the object's contour as described in detail in (*2*). In (*1*) the pixelation problem is circumvented by fitting the shape associated with a value of *E* to the contour data of the object. However, in every day use this is a time consuming and thus impractical method and a direct mapping of deformation to *E* is preferred.

The correct channel size is an issue of manufacturing and quality control. The specified range of deviation of the channel size contributes to the maximum error estimation of *E*.

This note will deal with the influence of pixelation for images with a pixel resolution of 340 nm/pix on deformation and size values as well as determining the viscosity values of *CellCarrier* and *CellCarrier B* in the flow-channel. Both media are based on standard 1x phosphate-buffered saline with the addition of long-chain methylcellulose polymers of about 0.5 w/v% and 0.6 w/v%, respectively. Such fluids with high viscosities due to polymer addition almost always display a shear-dependent rheology. In case of the *CellCarrier* media a shear-thinning effect is observed.

## A. Corrections for the influence of pixelation

To simulate the influence of pixelation circular objects were mapped on a pixel-grid with a random center position. For a pixel-accuracy contour, pixels were included as part of the object if the pixel's center position fell within the radius of the circle (Fig. 1).

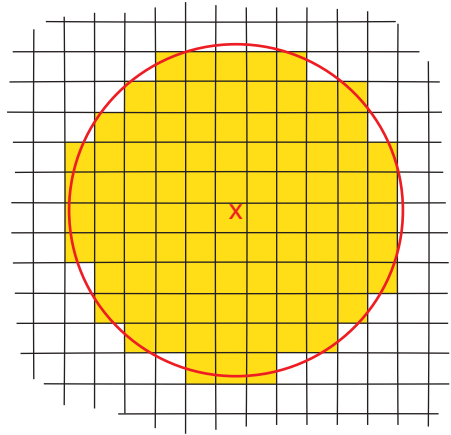

***Fig. 1. The pixelation effect.*** *A perfect circular outline positioned on a discrete pixel grid. Yellow-marked pixels belong to the circular disc object. The deformation of this object's pixel contour is larger than zero.*

For 99 different radii (between 0.4 µm and 20 µm) circular objects were simulated on a pixel grid with the pixel resolution of 340 nm/pix. At each radius 1000 random center positions were created and the obtained contours were analyzed in the same fashion as RT-DC data. A convex hull on the contour was used to calculate the size (as area inside the contour) and the deformation. The resulting deviations from the expected deformation value of zero are shown in Fig. 2.

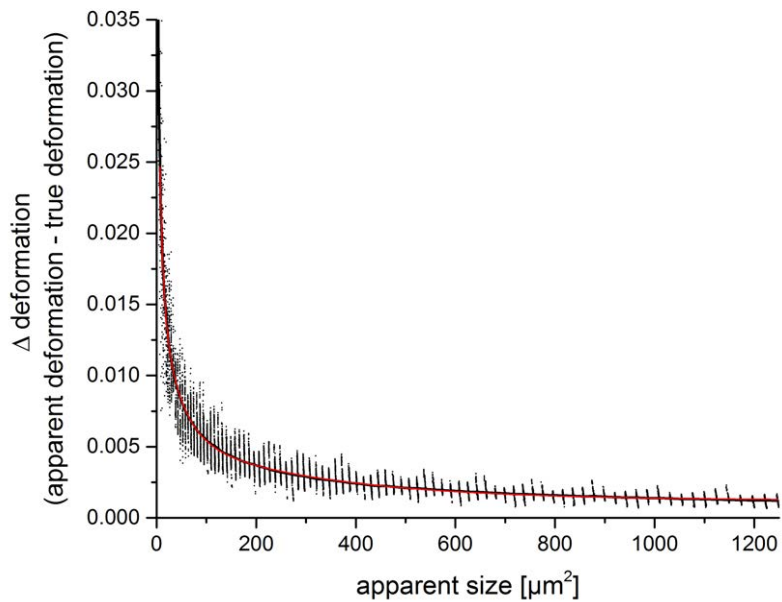

***Fig. 2. Deviation of apparent deformation from the theoretical value*** *depending on the object's size. For a pixel-grid resolution of 340 nm/pix, simulated deviations are shown as black dots. The apparent mean values for the different simulated sizes are shown as solid black line that is approximated by fitting a triple exponential decay (in red).*

The mean deformation deviation due to pixelation can be well approximated for a range of apparent sizes $A$ from 10 to 1250 μm² using a triple exponential decay:

(Eq. 1)

$$\Delta\, deformation = 0.0012 + 0.02 \cdot e^{-A/7.1} + 0.01 \cdot e^{-A/38.6} + 0.005 \cdot e^{-A/296}$$

with $A$ in μm².

In order to verify that the obtained size-dependent correction is valid for deformed objects as well, in a similar approach the pixelation-offsets were obtained for ellipses with a ratio of 0.75 and 0.5 between minor and major axis. The ratios of minor and major axis lead to theoretical, "true" deformation values of 0.0153 and 0.0828 respectively.

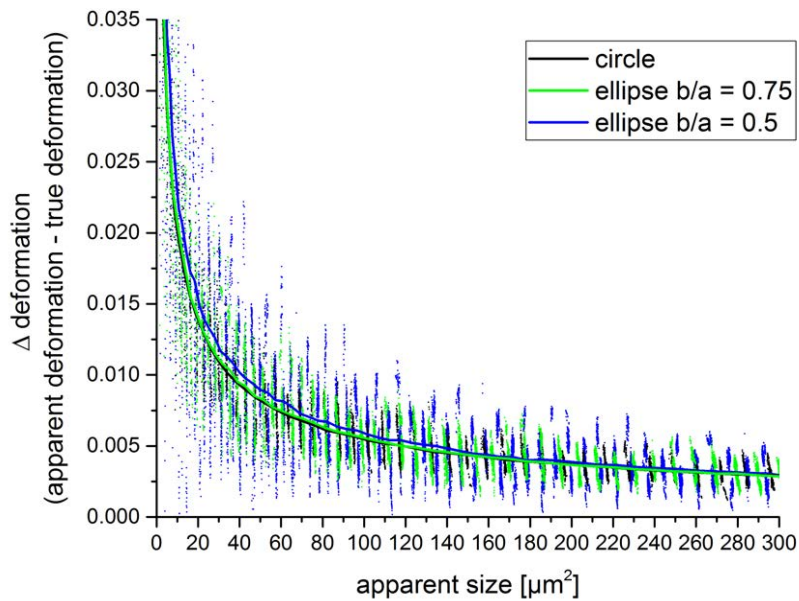

*Fig. 3. Deformation deviation due to pixelation for deformed objects.* For a pixel-grid resolution of 340 nm/pix, simulated deviations are shown for a circle (black dots) an ellipse with the ratio of minor to major axis of 0.75 (green dots, theoretical deformation 0.015) and an ellipse with the ratio of minor to major axis of 0.5 (blue dots, theoretical deformation 0.083). The apparent mean values for the different simulated object sizes are shown as solid lines.

The results are displayed in Fig. 3 and show a very similar behavior for objects with only slight deformations and large size – regardless of the deformation. At large deformations of small objects (< 100 μm²) the circular approximation underestimates the deformation offset by about 10 %, but can still serve as a reasonable correction for the mapping purpose.

While the deformation offset due to pixelation can be considered generally valid and directly applicable for mapping corrections, the corrections for deviations of the object size are more difficult. This is rooted in influences of the imaging setting, e.g. refractive index of the object, focal plane of the image, and gray-scale threshold for considering a pixel of the image part of the object. For the ideal binary case of the simulation setup, the apparent particle size seems to be underestimated compared to the theoretical size. This is shown by the ratio of true size and apparent size in Fig. 4.

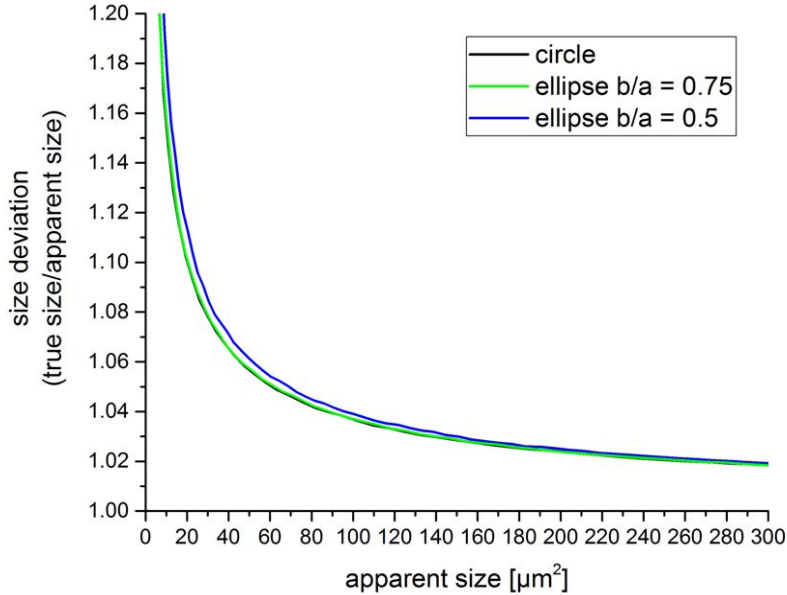

***Fig. 4. Size deviation due to pixelation for circular and deformed objects.*** *For a pixel-grid resolution of 340 nm/pix, simulated deviations are shown for a circle (black) an ellipse with the ratio of minor to major axis of 0.75 (green, theoretical deformation 0.015) and an ellipse with the ratio of minor to major axis of 0.5 (blue, theoretical deformation 0.083).*

The obtained size deviations are smaller than 10 % for objects larger 20 µm² and smaller than 5 % for objects larger 60 µm², which translates to diameter deviations of about 5 % at 5 µm and about 2.5 % at 9 µm.

In light of the comparably small size deviations, the dominating influence of the aforementioned experimental conditions and the lack of a suitable experimental verification – beads for possible calibration have manufacturing accuracies of the same magnitude as the simulated deviations – size corrections for pixelation in the mapping procedure seem unjustified.

To ease concerns on size corrections, Fig. 5 shows some examples of deviations in $E$ due to deformation and/or size deviations caused by pixelation for objects in a 20 µm channel. Size effects cause deviations of typically less than five percent and may therefore be neglected considering other experimental sources of error mentioned above. Deformation effects on the other hand can cause strong deviations in $E$, especially in the regime of low deformation. A correction for pixelation deformation allows one to reduce the major part of deviations in $E$ due to pixelation.

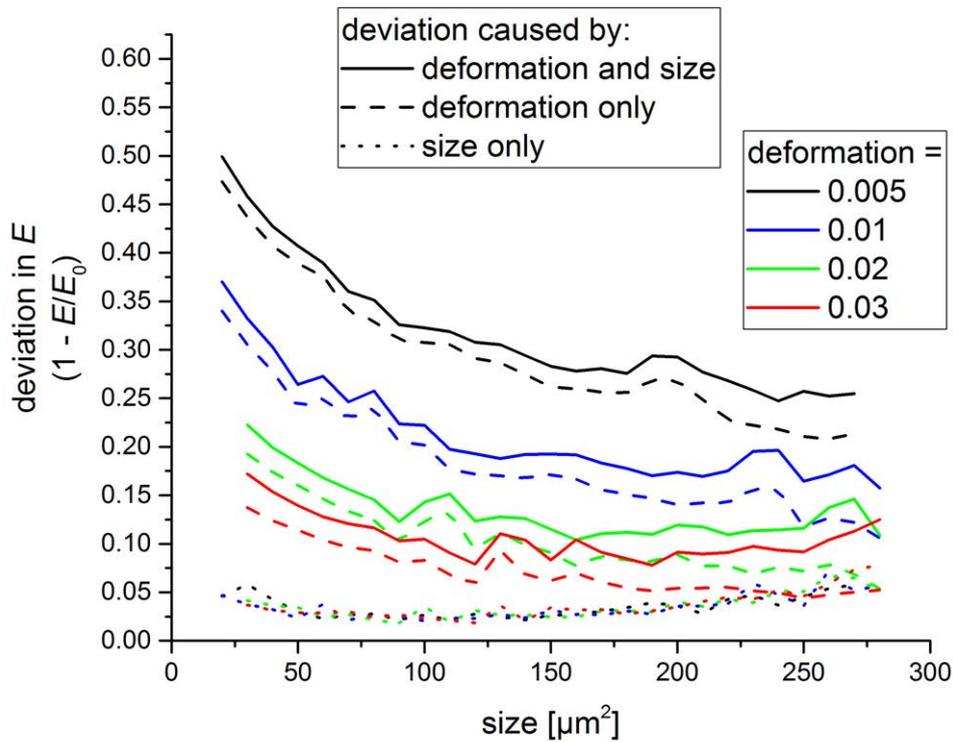

*Fig. 5. Possible deviations in E due to pixelation* for four elliptic shapes of different deformation in a 20 μm channel. The influence of size deviations only (dotted lines), deformation deviations only (dashed lines) and both size and deformation deviations (solid lines) is depicted.

In summary, for mapping deformation values to *E* a correction according to Eq. 1 should be employed. It yields results as a lower estimate for the mapped values of *E*.

**Note:** The suggested correction will work best for translating a population's center value (median or mean) of deformation to a value of *E*. For mapping of individual events in *E*, one should be aware that – especially for very small deformation – the lower half of the population's deformation distribution could become negative and the width of the distribution may in part be attributed to pixelation effects. Possible negative deformation values can either be neglected or mapped back to positive values by considering only absolute values. Both approaches lead to an overestimation of the deformation and therefore, an underestimation of *E*. These effects can be neglected for entire populations mapped in *E*, if all individual deformation values are above the correction value. A correction using deconvolution approaches could also be a remedy but is not considered for now.

## B. Finding the correct viscosity in the measurement channel

The viscosity of *CellCarrier* and *CellCarrier B* is controlled and tuned during production to yield 15 mPa s and 25 mPa s at 24 °C. Those values are obtained using a HAAKE falling ball viscometer and ball number 3. A shear-dependent behavior can already be observed using different balls yielding different viscosity values.

To further investigate the shear-dependent rheology, the dependence of shear stress $\tau$ on the shear rate $\dot{\gamma}$ needs to be determined.

The shear stress $\tau$ is proportional to the pressure difference $\Delta p$ over the flow channel and the shear rate $\dot{\gamma}$ is proportional to the flow rate $Q$. A Newtonian fluid, such as water, will yield a linear dependence between $\tau$ and $\dot{\gamma}$ and therefore also between $\Delta p$ and $Q$.

(Eq. 2)

$$\tau = \eta \cdot \dot{\gamma}$$

with the viscosity $\eta$.
Shear thinning is observed if the shear rate or flow rate grows faster than the shear stress or pressure difference. Shear thickening is observed if the shear rate grows slower than the shear stress.

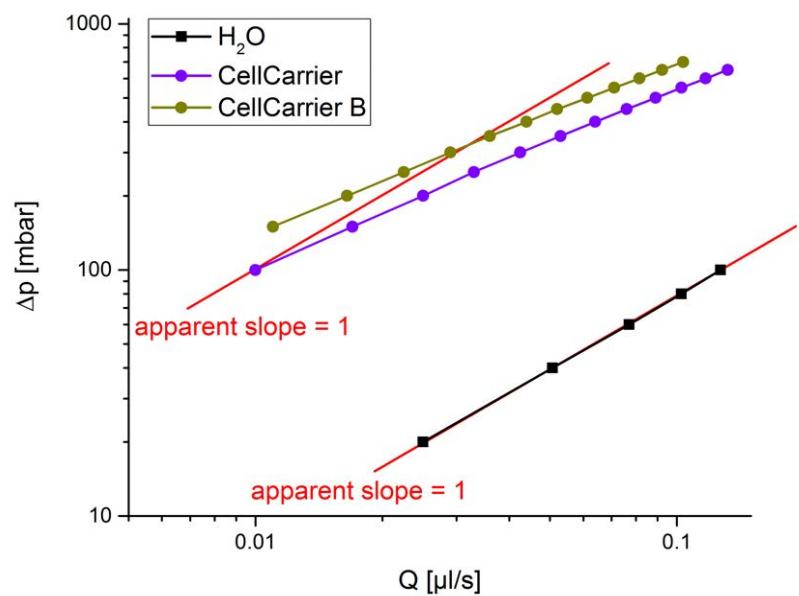

***Fig. 6. Relation between flow rate and pressure*** *for CellCarrier media and water in a Flic20. Red lines indicate an apparent slope of 1 equivalent with a linear dependence and the behavior of Newtonian fluids.*

Fig. 6 shows the dependence of flow rate (measured by Fluigent Flow-Rate Platform S) and pressure (controlled by Fluigent Flow Control) of water, *CellCarrier* and *CellCarrier B* at experimentally typical flow rates through a 20 µm Channel of a *Flic20*. Shear thinning behavior is observed for the *CellCarrier* media. A description of the shear thinning behavior by assuming a power law fluid seems applicable. The following relations describe a power law fluid:

(Eq. 3.1)
$$\tau = K \cdot \dot{\gamma}^n$$

and

(Eq. 3.2)
$$\eta = K \cdot \dot{\gamma}^{n-1}.$$

With knowledge about the exponent *n* and the amplitude *K* (that can be gathered from experimental determination) predictions for the viscosity of the *CellCarrier* media in the measurement channel can be made. They depend on the channel size and flow rate only. A good summary of relevant equations can be found in the introduction of (*3*).

For Newtonian fluids in a channel of length *L* with a squared cross-section of height and width *W*:

(Eq. 4.1)
$$\tau = \frac{W}{4} \frac{\Delta p}{L}$$

and

(Eq. 4.2)
$$\dot{\gamma} = 1.1856 \cdot \frac{6Q}{W^3}$$

For power law fluids in a squared cross-section channel:

(Eq. 5.1)
$$\tau = \frac{W}{4} \frac{\Delta p}{L}$$

and

(Eq. 5.2)
$$\dot{\gamma} = \left(1.1856 \cdot \frac{6Q}{W^3}\right) \cdot \frac{2}{3} \left(\frac{0.6771}{0.5928} + \frac{0.2121}{0.5928} \cdot \frac{1}{n}\right)$$

To calculate the viscosity of the *CellCarrier* in the measurement channel the pressure difference over the channel needs to be found. In order to consider solely the flow resistance of the measurement channel, inlet and outlet holes were punched directly before and after the channel. In addition, the pressure necessary to flow the fluids at a certain rate through the tubing and the flow rate measurement device was subtracted from the recorded pressure values.

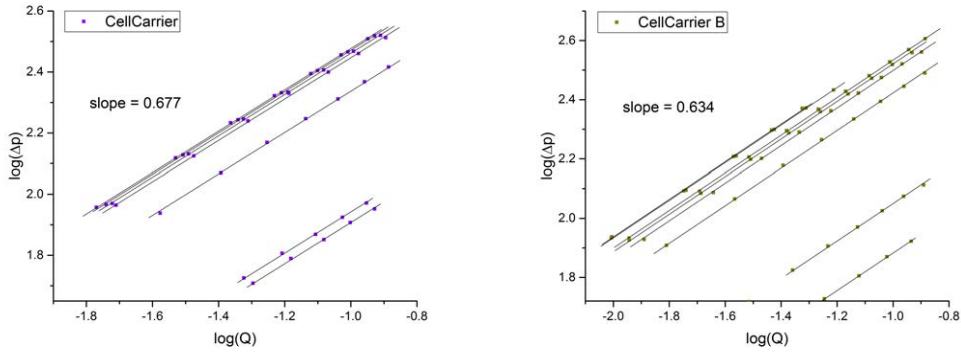

*Fig. 7. Slopes of pressure and flow rate in* **log-log** *space for CellCarrier (left panel) in five 20 μm and two 30 μm channels and CellCarrier B (right panel) in six 20 μm and two 30 μm channels at pressures and flow rates relevant for RT-DC experiments. The length of the channels varied depending on the inlet and outlet punching and was considered in the following calculation of the shear stress.*

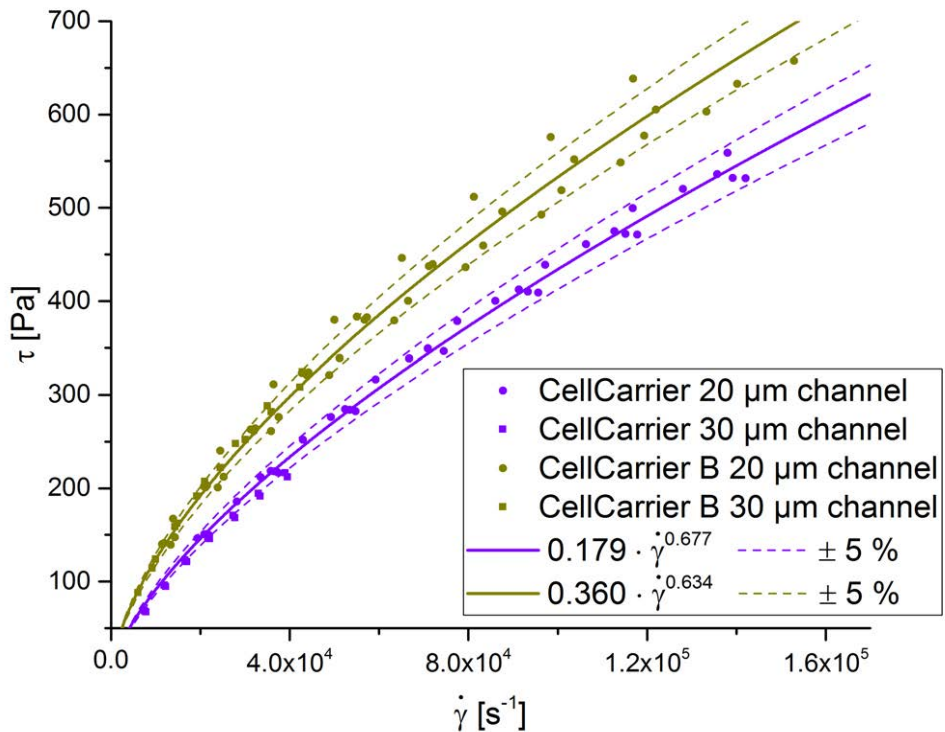

*Fig. 8. Shear rate vs. shear stress for CellCarrier media in 20 and 30 μm channels. Solid lines indicate best fits to all data. Dashed lines indicate the corridor of ± 5 % (of the viscosity value).*

The exponent *n* of the power law description can be determined from the slope between $\Delta p$ and $Q$ in log-log space (see Fig. 7). This yields $n = 0.677$ for *CellCarrier* and $n = 0.634$ for *CellCarrier B*.

With *n* known, the remaining task is to identify *K* in order to be able to calculate the correct viscosity in the channel. To this end, first the channel size was determined by recording the pressure - flow rate dependence for water in the same channel. Using Eq. 2, 4.1 and 4.2 the channel size $W$ (channel height and width) could be calculated since the viscosity of water is known and the individual channel length $L$ was obtained from microscopy image analysis. The resulting $\tau$ and $\dot{\gamma}$ were calculated according to Eq. 5.1 and 5.2 and are shown in Fig. 8 and yield coefficients $K = 0.179$ for *CellCarrier* and $K = 0.360$ for *CellCarrier B*.

The viscosity of the *CellCarrier* media – for a certain flow rate in a squared channel – can now be calculated according to Eq. 3.2 and 5.2, but an additional influence on the viscosity should be considered. The viscosity depends quite significantly on the temperature. Plotting the viscosity obtained by the falling ball viscometer at a temperature range of 18 °C to 26 °C in log-log space yields a power law dependence $\eta \sim \vartheta^{-0.866}$, with the temperature $\vartheta$ in °C (see Fig. 9).

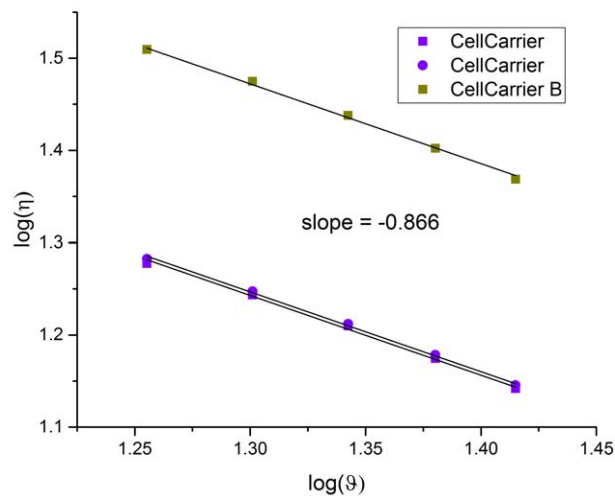

*Fig. 9. Temperature dependence of CellCarrier viscosity determined by falling ball viscometer experiments and plotted as logarithm. The recorded temperature range was 18 °C to 26 °C.*

With the addition of the temperature dependence the viscosity of *CellCarrier* is

(Eq. 6)

$$\eta = 0.179 \cdot \left(1.1856 \cdot \frac{6Q}{W^3} \cdot \frac{2}{3}\left(\frac{0.6771}{0.5928} + \frac{0.2121}{0.5928} \cdot \frac{1}{0.677}\right)\right)^{0.677-1} \cdot \left(\frac{\vartheta}{23.2}\right)^{-0.866}$$

and the viscosity of *CellCarrier B* is:

(Eq. 7)

$$\eta = 0.360 \cdot \left(1.1856 \cdot \frac{6Q}{W^3} \cdot \frac{2}{3}\left(\frac{0.6771}{0.5928} + \frac{0.2121}{0.5928} \cdot \frac{1}{0.634}\right)\right)^{0.634-1} \cdot \left(\frac{\vartheta}{23.6}\right)^{-0.866}$$

Table 1 shows a set of viscosities for typical flow rates in typical channel sizes.

| channel size $W$ [µm] | flow rate $Q$ [µl/s] | temperature $\vartheta$ [°C] | *CellCarrier* $\eta$ [mPa s] | *CellCarrier B* $\eta$ [mPa s] |
|---|---|---|---|---|
| 15 | 0.016 | 24 | 5.8 | 7.5 |
| 15 | 0.032 | 24 | 4.6 | 5.8 |
| 15 | 0.048 | 24 | 4.1 | 5.0 |
| 20 | 0.02 | 24 | 7.1 | 9.4 |
| 20 | 0.04 | 24 | 5.7 | 7.3 |
| 20 | 0.06 | 24 | 5.0 | 6.3 |
| 20 | 0.08 | 24 | 4.5 | 5.7 |
| 20 | 0.12 | 24 | 4.0 | 4.9 |
| 30 | 0.16 | 24 | 5.4 | 6.9 |
| 30 | 0.24 | 24 | 4.7 | 5.9 |
| 30 | 0.32 | 24 | 4.3 | 5.3 |
| 40 | 0.32 | 24 | 5.7 | 7.3 |
| 40 | 0.40 | 24 | 5.3 | 6.7 |
| 40 | 0.60 | 24 | 4.6 | 5.8 |

*Table 1. Viscosities at typical flow rates.*

In order to test the corrections, elastic hydrogel beads (diameter 14 µm) were measured in a 20 µm channel at different flow rates. The mapping of the obtained deformation values to apparent Young's moduli is shown with and without corrections in Fig. 10 and emphasizes the importance of the corrections made to obtain flow rate independent values of *E*.

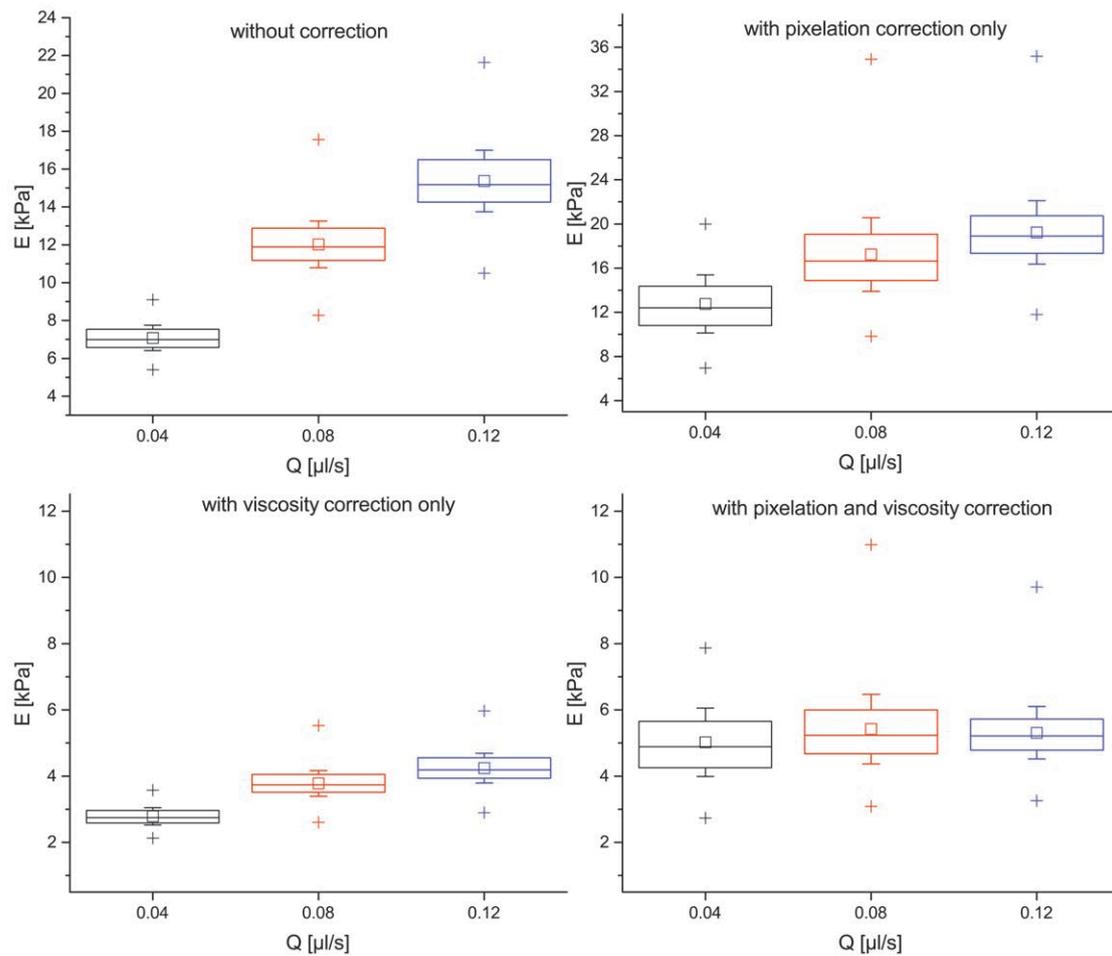

***Fig. 10. Influence of corrections on the apparent Young's Modulus*** *demonstrated by measurements of elastic hydrogel beads at different flow rates. Top left: mapping without any correction; top right: with pixelation correction only; bottom left: with viscosity correction only; bottom right: with viscosity and pixelation correction. Boxes indicate 1st quartile, median, 3rd quartile; squares indicate mean, whiskers indicate standard deviation and plus symbols indicate minimum and maximum.*

## References

1. A. Mietke, O. Otto, S. Girardo, P. Rosendahl, A. Taubenberger, S. Golfier, E. Ulbricht, S. Aland, J. Guck, E. Fischer-Friedrich, Extracting Cell Stiffness from Real-Time Deformability Cytometry: Theory and Experiment. *Biophys. J.* **109**, 2023–2036 (2015).

2. O. Otto, P. Rosendahl, A. Mietke, S. Golfier, C. Herold, D. Klaue, S. Girardo, S. Pagliara, A. Ekpenyong, A. Jacobi, M. Wobus, N. Töpfner, U. F. Keyser, J. O. R. Mansfeld, E. Fischer-Friedrich, J. Guck, real-time deformability cytometry: on-the-fly cell mechanical phenotyping. *Nat. Meth*ods **12**, 199–202 (2015).

3. Y. Son, Determination of shear viscosity and shear rate from pressure drop and flow rate relationship in a rectangular channel. *Polymer* **48**, 632–637 (2007).

## Acknowledgement

I want to thank Jochen Guck, Salvatore Girardo, Maik Herbig, Daniel Klaue and Philipp Rosendahl for support and discussions regarding this work.